\documentclass[twocolumn]{aastex6}
\pdfoutput=1 
\usepackage[T1]{fontenc}
\usepackage{amsmath,amstext}
\usepackage{apjfonts} 
\usepackage[figure,figure*]{hypcap}
\usepackage{microtype}

\begin{document}

\title{Radio Monitoring of the Tidal Disruption Event Swift J164449.3+573451. III. Late-time Jet Energetics and a Deviation from Equipartition }


\altaffiltext{1}{Harvard-Smithsonian Center for Astrophysics, Cambridge, MA 02138, USA}
\altaffiltext{2}{Center for Cosmology and Particle Physics, New York University, New York, NY 10003, USA}
\altaffiltext{3}{National Science Foundation, 2415 Eisenhower Avenue, Alexandria, VA 22314}
\altaffiltext{4}{Center for Interdisciplinary Exploration and Research in Astrophysics (CIERA) and Department of Physics and Astronomy, Northwestern University, Evanston, IL 60208, USA}

\author{T.~Eftekhari\altaffilmark{1}, 
E.~Berger\altaffilmark{1}, 
B.~A.~Zauderer\altaffilmark{2,3}, 
R.~Margutti\altaffilmark{4}, 
K.~D.~Alexander\altaffilmark{1}
}

\begin{abstract}

We present continued radio and X-ray observations of the relativistic tidal disruption event Swift J164449.3+573451 extending to $\delta t \approx 2000$ d after discovery. The radio data were obtained with the VLA as part of a long-term program to monitor the energy and dynamical evolution of the relativistic jet and to characterize the parsec-scale environment around a previously dormant supermassive black hole. We combine these data with $\textit{Chandra}$ X-ray observations and demonstrate that the X-ray emission following the sharp decline at  $\delta t \approx 500$ d is due to the forward shock. Using the X-ray data, in conjunction with optical/NIR data, we constrain the synchrotron cooling frequency and the microphysical properties of the outflow for the first time. We find that the cooling frequency evolves through the optical/NIR band at $\delta t \approx 10 - 200$ d, corresponding to a magnetic field energy density fraction of $\epsilon_B \approx 10^{-3}$, well below equipartition; the X-ray data demonstrate that this deviation from equipartition holds to at least $\delta t \approx 2000$ d. We thus recalculate the physical properties of the jet over the lifetime of the event, no longer assuming equipartition. We find a total kinetic energy of $E_K \approx 4 \times 10^{51}$ erg and a transition to non-relativistic expansion on the timescale of our latest observations ($\delta t \approx 700$ d). The density profile is approximately $R^{-3/2}$ at $\lesssim 0.3$ pc and $\gtrsim 0.7$ pc, with a plateau at intermediate scales, characteristic of Bondi accretion. Based on its evolution thus far, we predict that Sw 1644+57 will be detectable at centimeter wavelengths for decades to centuries with existing and upcoming radio facilities. Similar off-axis events should be detectable to $z \sim 2$, but with a slow evolution that may inhibit their recognition as transient events.
\end{abstract}

\bigskip


\section{Introduction}\label{sec:intro}

The unusual $\gamma$-ray/X-ray transient Swift J164449.3+573451 (hereafter, Sw 1644+57; $z=0.354$) marked a significant departure from the standard tidal disruption scenario \citep{Rees1988} as the first such event leading to a relativistic jet (\citealt{Bloom2011}; \citealt{Burrows2011}; \citealt{Levan2011}; \citealt{Zauderer2011}). This rare event provides an unprecedented opportunity to trace the complete evolution of a relativistic jet and to characterize the parsec-scale environment around a previously-dormant supermassive black hole (SMBH). Radio observations initiated within hours of the $\gamma$-ray trigger revealed a brightening radio source, coincident with the nucleus of the host galaxy and resulting from a relativistic jet launched at the time of disruption \citep{Zauderer2011}. A tidal disruption origin was further supported by rapid variability of the X-ray emission ($\lesssim$ 100 s), corresponding to a central engine a few times the Schwarzschild radius of a $\sim 10^6 \rm \ M_{\odot}$ black hole, and a mean X-ray luminosity of $\sim 10^{47} \rm \ erg \ s^{-1}$, roughly $2-3$ orders of magnitude above the Eddington limit of the black hole \citep{Burrows2011}.

In recent years, the number of observed tidal disruption events (TDEs) has risen drastically. To date, there are about 60 TDE candidates cited in the literature,\footnote{See https://tde.space} detected primarily in optical and X-rays (see \citealt{Komossa2015}, \citealt{Kochanek2016}, \citealt{Auchettl2017} for recent reviews). Although these events share a number of characteristic similarities, their apparent diversity indicates that the TDE phenomenon is quite complex, with gas outflows perhaps playing a key role in the observable signatures (e.g., \citealt{Chornock2014}, \citealt{Miller2015}, \citealt{Alexander2016}). In this context, continued observations of Sw 1644+57 provide a unique opportunity to enhance our understanding of these rare events, with additional implications for other high-energy astrophysical phenomena, including gamma-ray bursts (GRBs) and active galactic nuclei (AGN). 

In the years since its discovery, observations across the electromagnetic spectrum have traced the evolution of Sw 1644+57. \citet{Zauderer2013} present X-ray observations extending to $\delta t \approx 600$ d, which revealed a sharp decline by more than two orders of magnitude at $\delta t \approx 500$ d (see also \citealt{Mangano2016} and \citealt{Levan2016} where optical and infrared data spanning a year after the discovery of the event are presented). We have been carrying out a long-term monitoring campaign of the radio emission from Sw 1644+57. Our previous work has tracked the expanding outflow from Sw 1644+57, with observations extending to $\delta t \approx$ 600 d presented in \citet{Zauderer2011} (hereafter, ZBS11), \citet{Berger2012} (hereafter, Paper I) and \citet{Zauderer2013} (hereafter, Paper II). This work has led to a number of critical insights:

\begin{itemize}

\item{The discovery of bright synchrotron emission ($F_\nu \approx 35$ mJy) with an initial peak in the millimeter band and a steep, self-absorbed power law at lower frequencies, revealing the presence of a relativistic outflow with $\Gamma \sim$ few, and a launch date consistent with the time of disruption (ZBS11; see also \citealt{Bloom2011}).}

\item{An order of magnitude increase in the jet energy at $\delta t \approx 30 - 250$ d, indicative of a structured outflow launched with a range of Lorentz factors, $E(>\Gamma_j)\propto \Gamma_j^{-2.5}$ (Paper I).}

\item{A sharp decline in the X-ray flux by a factor of 170 at $\delta t \approx 500 \rm \ d$ as the relativistic jet shut off. Assuming this timescale corresponds to $\dot{M}\approx \dot{M}_{\rm Edd}$, this led to an inferred accreted mass of $\approx 0.15 \rm \ M_{\odot}$, consistent with the tidal disruption of a $\lesssim 1 \rm \ M_{\odot}$ star (Paper II).}

\item{An overall density profile around the SMBH of $\rho \propto R^{-3/2}$ between $0.1-0.2$ pc and a flattening between $0.4 - 0.6$ pc, consistent with the expected Bondi radius for a $10^6 - 10^7 \rm \ M_{\odot}$ black hole (Paper I).}
\end{itemize}

In this work, we present new radio and X-ray observations extending to $\delta t \approx 2000$ d providing a detailed view of the emission when the X-ray flux is in a low state, and the jet transitions to non-relativistic expansion. We further combine the new and previous radio and X-ray observations with the optical/NIR data \citep{Levan2016} to explore for the first time the forward shock emission across nine orders of magnitude in frequency and three orders of magnitude in time.

The paper is organized as follows. We describe the radio and X-ray observations in  \S\ref{sec:obs}. A detailed explanation of our modeling of the radio emission is presented in \S\ref{sec:model}. In  \S\ref{sec:results}, we utilize the data and model to track the jet energy and Lorentz factor, to determine the shock microphysical parameters, and to determine the overall density profile of the ambient medium. We also explore alternative origins of the late-time X-ray emission. Finally, we consider the implications of our results in the context of off-axis TDEs in \S\ref{sec:offaxis}.

\section{Observations}\label{sec:obs}

\subsection{Radio Observations}\label{sec:radioobs}

Our previous work has tracked the radio evolution of Sw 1644+57 to $\delta t \approx$ 580 days (ZBS11, Paper I, Paper II). In this work, we report observations spanning $\delta t \approx 600-1900$ days. Although the event was first triggered by the \textit{Swift} Burst Alert Telescope (BAT) on  2011 March 28.55 UT, subsequent analysis of the BAT data revealed discernible emission as early as 2011 March 25 \citep{Burrows2011}. All times discussed here are therefore relative to 2011 March 25.5 UT. 

We obtained radio observations with the Karl G. Jansky Very large Array (VLA\footnote{The VLA is operated by the National Radio Astronomy Observatory, a facility of the National Science Foundation operated under cooperative agreement by Associated Universities, Inc. The observations presented here were obtained as part of programs 13A-049, 14A-423 and 15B-273.}). We performed bandpass and flux density calibration using 3C286 for all frequencies and epochs. For phase calibration in L ($\sim$1.5 GHz) and S ($\sim$3 GHz) bands, we used J1634+6245, while at all other frequencies we used J1638+5720. We processed and imaged the data using the Common Astronomy Software Application (CASA) software package \citep{McMullin2007}. The flux densities and corresponding uncertainties were measured using the \texttt{imtool} program as part of the \texttt{pwkit}\footnote{Available at https://github.com/pkgw/pwkit} package \citep{pwkit} (see Table 1).

\subsection{X-Ray Observations}\label{sec:xrayobs}
We obtained X-ray observations with the \textit{Chandra Advanced CCD Imaging Spectrometer} (ACIS-S) (PI: Zauderer). The observations were obtained in five epochs spanning $\delta t \approx 1610-2020$ days, with exposure times of 19.7$-$26.7 ks; see Table 2. We analyzed the data using $\texttt{CALDB}$ within the $\texttt{CIAO}$ software package (v4.7) following standard ACIS data filtering. For source detection, we perform a targeted extraction using an aperture size of $1.5''$ and an annular background region with inner and outer radii of $2''$ and $6''$, respectively, obtaining an individual net count rate and source significance for each observation. For uniformity, we also reanalyze with the same procedure {\it Chandra} observations from 2012 and 2015 (\citealt{Levan2012}; \citealt{Zauderer2013};  \citealt{Levan2016}).

We find a relatively low number of counts at each epoch ($\sim 2-13$), including a non-detection at $\delta t \approx 1611$ d. Thus, we perform a simultaneous fit to the data by fixing the fit parameters across all observations. We use an absorbed power law spectrum (\texttt{tbabs*ztbabs*pow} in \texttt{xspec}) with a Galactic absorption of $N_{\rm H,MW}=1.7\times10^{20} \ \rm{cm^{-2}}$ (\citealt{Willingale2013}; see also \citealt{Kalberla2005}) and an intrinsic absorption column of $N_{\rm H,int}=2\times10^{22} \ \rm{cm^{-2}}$ \citep{Mangano2016}. For the purpose of constraining the power law index, $\Gamma_X$, we use the six epochs spanning $2015-2016$, excluding the 2015 non-detection, and find $\Gamma_X=1.66\substack{+0.72 \\ -0.66}$ (1$\sigma$). 

For forward shock emission with $\nu_X > \nu_c$ and $p=2.5$ (see \S\ref{subsec:seds}), we expect $\Gamma_X = 2.25$  ($\Gamma_X=p/2+1$), which is consistent with the measured value within 1$\sigma$. We therefore impose $\Gamma_X=2.25$ to calculate the unabsorbed fluxes (3-10 keV) for each individual epoch (Table 2). We note that a number of alternative explanations for the residual X-ray flux beginning at $\delta t \approx 500$ d can be ruled out. We explore each of these scenarios in \S\ref{sec:latexray}.

\section{Modeling of the Radio Emission}\label{sec:model}

\subsection{Spectral Energy Distributions}\label{subsec:seds}

We model the individual radio spectral energy distributions (SEDs) to constrain the temporal evolution of the outflow and the radial profile of the ambient medium (see Figure 1). We use the prescription for GRB afterglows \citep{Granot2002}, in which the observed radiation is due to non-thermal synchrotron emission. We assume that electrons in the shock are accelerated into a power law distribution, $N(\gamma_e) \propto \gamma_e^{-p}$ for $\gamma_e > \gamma_m$, where $\gamma_m$ is the minimum Lorentz factor. The SED  is characterized by a peak specific flux density, $F_{\nu,p}$ at a frequency $\nu_p$. In the slow cooling regime, $v_p$ corresponds to the self-absorption frequency ($\nu_a$), or the frequency corresponding to $\gamma_m$ ($\nu_m$).

 At early times, when $\nu_a < \nu_m$, the synchrotron spectrum is given by \citep{Granot2002}:

\begin{multline}
\indent F_1 \equiv F_\nu (\nu_a) \Bigg[ \Big(\frac{\nu}{\nu_a}\Big)^{-s_1\beta_1}+\Big(\frac{\nu}{\nu_a}\Big)^{-s_1\beta_2}\Bigg] \ \times 
\\ 
\Bigg[1+\Big(\frac{\nu}{\nu_m}\Big)^{-s_2 (\beta_2-\beta_3)}\Bigg]^{-1/s_2},
\end{multline}

\noindent where $\beta_n$ describes the spectral slopes above and below a break [here, $\beta_1 = 2$, $\beta_2=1/3$, $\beta_3 = (1-p)/2$], and $s_1$ and $s_2$ are smoothing parameters that describe the shape of the spectrum across each break. Following our previous findings in ZBS11, Paper I, and Paper II, we adopt $p$ = 2.5. The hydrodynamical evolution of the shock is adiabatic and $\nu_m$ therefore decreases more rapidly with time than $\nu_a$  \citep{Sari1997}, leading to $\nu_m < \nu_a$ at late times. The spectrum in this regime is given by \citep{Granot2002}:

\begin{multline}
\indent F_2 \equiv F_\nu (\nu_m) \Bigg[ \Big(\frac{\nu}{\nu_m}\Big)^2 e^{-s (\frac{\nu}{\nu_m})^{2/3}}+\phi^{5/2}\Bigg] \ 
\times \\
\Bigg[1+\Big(\frac{\nu}{\nu_a}\Big)^{-s_2 (\beta_2-\beta_3)}\Bigg]^{-1/s_2},
\end{multline}

\noindent where $\beta_2 = 5/2$ and $\beta_3 = (1-p)/2$. In  \S\ref{sec:coolfreq}, we show that the cooling frequency ($\nu_c$) is located well above $\nu_p$ and therefore does not affect the radio SEDs. 

The shape of each power-law segment is calculated in the asymptotic limit. The above spectra are therefore well-defined when the break frequencies are well separated. As the break frequencies evolve and the SED transitions from being described by Equation 1 to Equation 2, neither equation is strictly valid; we therefore implement a weighting scheme which allows for a smooth evolution across the two regimes:

\begin{equation}
F = \frac{w_1F_1 + w_2F_2}{w_1 + w_2}
\end{equation}

\noindent where $w_1 = (\nu_m/\nu_a)^2$ and $w_2 = (\nu_a/\nu_m)^2$. We fit this model to all radio SEDs to extract $\nu_a$, $\nu_m$ and $F_{\nu, p}$ at each epoch. We show the resulting time evolution of $\nu_a$, $\nu_m$, and $F_{\nu, p}$ in Figure 2. We find that at $\delta t \approx 300 - 400$ d, the spectrum transitions into Equation 2 as $\nu_m$ decreases below $\nu_a$. This transition is accompanied by a decrease in the peak flux.

\subsection{Equipartition Analysis}\label{subsec:equip}

Using the extracted values of $\nu_a$, $\nu_m$ and $F_{\nu, p}$ from the SED fitting, we infer the properties of the blastwave and local environment (see Figures 3 and 4). The minimum energy of the blastwave can be derived by assuming that the electron and magnetic field energy densities are in equipartition (\citealt{Pacholczyk1970}; \citealt{Scott1977}; \citealt{Chevalier1998}; \citealt{BarniolDuran2013}). These depend on the size of the emitting region, providing a robust estimate of the source radius. The system can be further characterized by the bulk Lorentz factor ($\Gamma$), the co-moving magnetic field strength ($B$), and the total number of radiating electrons in the emitting region ($N_e$). The equipartition energy and radius are given by \citep{BarniolDuran2013}:

\begin{equation} 
E_{\rm eq} \approx (2.5\times10^{49} \textrm{erg}) \Bigg[\frac{F_{p,\rm mJy}^{\frac{20}{17}} d_{L,28}^{\frac{40}{17}}  \eta^{\frac{15}{17}}} {\nu_{p,10} (1+z)^{\frac{37}{17}}}\Bigg] \frac{f_{V}^{\frac{6}{17}}}{\Gamma^{\frac{26}{17}} f_{A}^{\frac{9}{17}}}
 \end{equation}

\begin{equation} 
R_{\rm eq} \approx (1.7\times10^{17} \textrm{cm}) \Bigg[\frac{F_{p,\rm mJy}^{\frac{8}{17}} d_{L,28}^{\frac{16}{17}} \eta^{\frac{35}{51}}} {\nu_{p,10} (1+z)^{\frac{25}{17}}}\Bigg]
\frac{ \Gamma^{\frac{10}{17}}}{ f_A^{\frac{7}{17}} f_V^{\frac{1}{17}}}
 \end{equation}

\noindent where $d_L$ is the luminosity distance, $\eta = \nu_p/\nu_a$ for $\nu_m > \nu_a$ and $\eta = 1$ for $\nu_m < \nu_a$, $f_A$ and $f_V$ are geometric filling factors which parameterize deviations from a spherical geometry and describe the fraction of the observed area and volume subtended by the source, and we use the notation $X \equiv10^yX_y$. In the case of Sw 1644+57, the geometry is best described by a ``narrow" jet, in which the jet half-opening angle $\theta_j$ (assumed to be $\sim$ 0.1 as in our previous work) is less than the relativistic beaming angle $1/\Gamma$. The observer therefore sees the entirety of the jet beamed into a region larger than the jet itself. The filling factors in this scenario are given by $f_A$ = $f_V$ = $(\theta_j \Gamma)^2$. As in \citet{BarniolDuran2013}, $\Gamma$ can be obtained by relating the radius to the time since the onset of the outflow, using $\delta t=R(1-\beta)(1+z)/(\beta c)$. The values of $B$ and $N_e$ are similarly obtained using the equations given in \citet{BarniolDuran2013}. The number density in the surrounding ambient medium can be calculated using $n_{\rm ext} = n_e/4\Gamma^2$, where $n_e =N_e/V$ is the number density of radiating electrons in the outflow, and $V$ is the volume of the emission region \citep{BarniolDuran2013}. 

As in our previous work, we assume that the fraction of post-shock energy in the electrons is $\epsilon_e$ = 0.1. If the kinetic energy of the outflow is dominated by protons, the equipartition energy will increase by a factor of $\xi^{11/17}$ and the radius will increase by a factor of $\xi^{1/17}$, where $\xi = 1 + \epsilon_e^{-1} \approx 11$. Moreover, any deviation from equipartition is parameterized by the factor $\epsilon = (\epsilon_B/\epsilon_e)(11/6)$, which introduces a multiplicative factor of $\epsilon^{1/17}$ for the radius and $(11/17)\epsilon^{-6/17} + (11/17)\epsilon^{7/12}$ for the energy. 

When $\nu_m < \nu_a$, the bulk of the electron energy is in electrons whose radiation is self-absorbed. This requires a correction factor of $(\nu_a/\nu_m)^\frac{p-2}{2}$ to the value of $N_e$. As a result, the energy is increased by a factor $(\nu_m/\nu_a)^{\frac{11(2-p)}{34}}$, and the radius is increased by a factor $(\nu_m/\nu_a)^{\frac{(2-p)}{34}}$. These correction factors subsequently propagate through to calculations of the other parameters.

We note that our treatment in this work is consistent with that presented in \citet{BarniolDuran2013b}, in which the authors use the results of our Papers I and II in the context of an equipartition formulation to derive the physical properties of the outflow. Here, we repeat the same analysis, using the results from the SED fits presented in this work and imposing our constraints on $\epsilon_B$. Finally, we note that in the non-relativistic regime ($\Gamma \beta \approx 1$), the mildly-relativistic formalism still holds with $\Gamma \approx 1$; see Equations 21 and 25 of \citet{BarniolDuran2013}.

\subsection{MCMC Fitting}\label{subsec:model}

Our previous modeling of the radio data (ZBS11, Paper I, Paper II) utilized frequentist methods and therefore lacked robust uncertainties on the estimated physical parameters that characterize the outflow and ambient medium. In this work, we implement Parallel-Tempered Markov Chain Monte Carlo (MCMC) methods to fit the broadband radio SED at each epoch using the Python-based implementation \texttt{emcee} \citep{ForemanMackey2013}. Parallel tempering utilizes multiple-chains at unique temperatures to efficiently explore multi-modal posterior distributions \citep{Geyer1991}. We use a simple Gaussian likelihood with model parameters $\nu_a$, $\nu_m$, and $F_{\nu}$ (Equations $1-3$), and an additional parameter which accounts for underestimates of the uncertainties on individual data points. We use linearly uniform priors, where the upper bounds for the break frequencies are iteratively informed by the results from each previous epoch given the expectation of smooth time evolution from high to low frequencies as the blastwave decelerates. We report lower limits for $\nu_m$ for several epochs where the data do not span the peak of the SED. The posterior distributions are sampled using 100 Markov chains. We use a total of 10 temperature modes and report the parameter estimates corresponding to the lowest temperature, or unmodified posterior. Auto-correlation lengths are typically of the order of $\sim$90 steps. To ensure that our samples represent independent, uncorrelated measures of the target distribution, we run each Markov chain for 3000 steps, discarding the first 500 steps and ensuring that the samples have sufficiently converged beyond this point. As a final metric to assess the quality of our MCMC samplers, we calculate the acceptance fraction, the fraction of proposed steps that are accepted in a given chain. We find that the average acceptance fraction spans $0.34-0.55$ for all of our fits, within the nominally accepted efficiency rate \citep{Roberts2001}. The posterior distributions from the MCMC analysis are used to calculate corresponding uncertainties for the physical parameters described in \S\ref{subsec:equip}.

\section{Results}\label{sec:results}
\subsection{Constraining the cooling frequency and $\epsilon_B$}\label{sec:coolfreq}

In the slow cooling regime, only the highest energy electrons which populate the tail-end of the distribution can cool efficiently. These electrons emit above the cooling frequency ($\nu_c$) with a Lorentz factor of $\gamma_c$, which depends inversely on $\epsilon_B$, $\gamma_c \propto 1/\epsilon_B$ \citep{Sari1998}. 
The signature of synchrotron cooling is an additional spectral break at $\nu_c$ with $F_\nu \propto \nu^{-p/2}$ above the break. The cooling frequency is given by \citep{Sari1998}:

\begin{equation} 
\nu_c = 2.8\times10^{6} \gamma_c^2 \Gamma^2 B
 \end{equation}

\noindent where $\gamma_c = 5\times10^9 \epsilon_B^{-1} t_d^{-1} n_{\rm ext}^{-1} \Gamma^6$. 

We search for the cooling break using the host-subtracted optical/NIR measurements from \citet{Levan2016}. Specifically, we extend the fits of our individual radio SEDs to the optical/NIR regime and quantify the effect of the cooling break (see Figure 5). When $\nu_c$ is located below the optical/NIR regime at all times (i.e., when $\epsilon_B \gtrsim$ 0.1, or essentially in equipartition), the extrapolations underpredict the observed flux densities (e.g., by a factor of 5 in K band). Since $\nu_c \propto \epsilon_B^{-2}$, decreasing $\epsilon_B$ shifts $\nu_c$ above the optical/NIR regime resulting in extrapolations that eventually overpredict the observed flux densities. As the optical/NIR flux measurements represent only the transient source flux, any excess flux in the extrapolated values relative to the observed measurements is therefore indicative of host galaxy extinction, already noted to be present in previous analyses (ZBS11, \citealt{Levan2011}).  We determine the rest-frame extinction, $A_i$, assuming a Milky Way extinction curve with $R_v=3.1$ (\citealt{Fitzpatrick1999}; \citealt{Draine2003}). Though largely negligible, we also include contributions from Galactic extinction \citep{Schlafly2011}. We conclude that $\epsilon_B \approx 10^{-3}$ with $A_i \approx 3$ mag, corresponding to a cooling frequency that transitions through the optical/NIR over the timescale $\delta t \approx 10-200$ d. We note that a lower value of $\epsilon_B$ can also be accommodated by the optical/NIR data, provided that the extinction is even larger (e.g., $\epsilon_B \approx 10^{-5}$ and $A_i \approx 4.5$ mag). However, as we show below this leads to a disagreement with the X-ray light curve.

The overall evolution of the radio SEDs predicts the optical/NIR light curves remarkably well, indicating that the optical/NIR emission is due to the  same blastwave that produces the radio emission. The apparent re-brightening, or "optical bump", seen at $\delta t \approx 30$ d can therefore be explained within the context of the relativistic outflow, as suggested in \citet{Levan2016}. Namely, the re-brightening seen in the optical is a consequence of the structured outflow which leads to an increase in energy as slower moving matter catches up with the decelerated ejecta (Paper I; see also \citealt{deColle2012}). 

We similarly find that the X-ray emission beginning at $\delta t \approx 500$ d originates in the forward shock, and show that it can also be explained with $\epsilon_B\approx 10^{-3}$ (see Figure 6). At $\delta t \approx 500$ d, an abrupt drop in the X-ray luminosity revealed a shift in the emission mechanism (Paper II). The residual X-ray flux is too bright for direct thermal emission from the accretion disk, which led us to suggest a forward shock origin (Paper II; we consider alternative scenarios for the X-ray emission in \S\ref{sec:latexray}). The X-ray observations presented here provide a much longer time baseline over which we can test this model. As with the optical/NIR data, we extrapolate the radio SEDs into the X-ray regime  and compare to the measured X-ray fluxes (see \S\ref{sec:xrayobs}). We find that assuming equipartition ($\epsilon_B \approx$ 0.1) leads to an underestimate of the observed X-ray fluxes by about an order of magnitude. On the other hand, $\epsilon_B \approx 10^{-3}$ provides an excellent match to the X-ray data over the period $\delta t \approx 500 - 2000$ d. 

Although the optical/NIR data can be accommodated with a much lower value of $\epsilon_B$ ($\approx 10^{-5}$), this corresponds to a cooling frequency that transitions through the X-ray regime, resulting in extrapolations which overpredict the observed X-ray fluxes by roughly an order of magnitude. Thus, we find that $\epsilon_B\approx 10^{-3}$, coupled with the forward shock SED evolution captured in the radio band, explain both the optical/NIR light curves at $\delta t\approx 10-200$ d and the X-ray light curve at $\delta t\approx 500-2000$ d.  The cooling frequency evolves from the NIR to the optical at $\delta t\approx 10-200$ d, and is located at $\approx 10^{16}$ Hz at $\delta t\approx 1900$ d.

\citet{BarniolDuran2013b} consider an outflow that approaches equipartition at late times. In particular, they invoke a time-varying deviation from equipartition to explain the apparent increase in energy between $\delta t \approx 30-250$ d. This scenario avoids late-time energy injection, suggesting that the total energy is initially launched at the outset and simply converted from one form to the other (via magnetic reconnection and inverse Compton, for example). However, we find that a constant value of $\epsilon_B$ fits the entire optical/NIR light curves and the late-time X-ray emission remarkably well, spanning $10-1900$ d.  This argues against the scenario of \citet{BarniolDuran2013b} and instead lends continued credence to our suggestion of energy injection from a structured Lorentz factor profile (Paper I). We note that inverse-Compton (IC) cooling of synchrotron electrons by X-ray photons has also been invoked to explain the increase in energy \citep{Kumar2013}. This model relaxes the assumption that $\epsilon_B$ should vary in time, and instead suggests that IC cooling suppresses the emission from synchrotron electrons at early times. However, because the effects of IC cooling modify the synchrotron spectrum, a test of this model is beyond the scope of this work.

\subsection{Alternative Origins of the Late-time X-ray Emission?}\label{sec:latexray}

The rapid decline in the X-ray flux at $\delta t \approx 500$ d suggests a fundamental change in the nature of the emission. In \S\ref{sec:coolfreq}, we demonstrate that the X-ray flux between $\delta t \approx 500 - 2000$ d can be explained by synchrotron emission from the forward shock. Here we consider a number of alternative explanations, including thermal emission from the accretion disk, X-ray emission due to star formation, or the presence of an AGN. 

In Paper II, we explored the possibility that the residual X-ray flux beginning at $\delta t \approx 500$ d is due to thermal emission from the TDE accretion disk itself. In this model, the expected disk temperature ($\sim 30-60$ eV, at the critical accretion rate $\sim \dot M_{\rm Edd}$) at the inner radius for a $\approx 10^6 \ \rm M_{\odot}$ black hole corresponds to an X-ray flux $\lesssim 10^{-16} \rm \ erg \ cm^{-2} \ s^{-1}$ (\citealt{Mangano2016}; Paper II), well below the observed flux. Moreover, the flux normalization in the spectrum requires an inner disk temperature of $\approx 1$ keV, well in excess of the temperatures appropriate for a $\sim 10^6 - 10^7 \rm \ M_{\odot}$ black hole (Paper I). A self-consistent thermal disk model is therefore unable to reproduce the observed late-time X-ray emission. 

We can similarly rule out X-ray emission due to star formation. We estimate the expected star formation rate (SFR) corresponding to an observed X-ray luminosity of $L_X \approx 3 \times 10^{42} \rm \ erg \ s^{-1}$ \citep{Grimm2003}, and find SFR $\approx 70 \rm \ M_{\odot} \ yr^{-1}$, in strong disagreement with the value of $0.5 \rm \ M_{\odot} \ yr^{-1}$, as estimated for the host galaxy of Sw 1644+57 using emission lines \citep{Levan2011}. We note that in the case of highly obscured star formation due to the absorption of Lyman continuum photons, the strength of the optical emission lines may be attenuated by some non-negligible factor, resulting in underestimates of the derived star formation rate (\citealt{Inoue2001}; \citealt{Dopita2003}). However, this effect, which scales with the dust-to-gas ratio, introduces at most a correction factor of $2-5$ to the estimated star formation rate \citep{Inoue2001}, and therefore cannot be reconciled with the rate inferred from the X-ray emission.

Finally, we argue against an AGN origin for the late-time X-ray emission on the basis of the observed steady temporal decline in the X-ray flux. Fitting a linear function to the X-ray data points over the period $\delta t \approx 500-2000$ d, we find that the flux decreases with time with a slope of $-0.3$. A steady decline is expected in the forward shock scenario but is not obviously expected for an AGN. Furthermore, if the observed X-ray emission beginning at $\delta t \approx 500$ d was due to an AGN, this would place a limit on the forward shock contribution of  $\lesssim 10^{-15}  \rm \ erg \ s^{-1} \ cm^{-2}$. However, this in turn requires $\epsilon_B \gtrsim 10^{-3}$, in contradiction with the inference from the optical/NIR data (see \S\ref{sec:coolfreq}). In \citet{Kathirgamaraju2017}, the authors suggest that the sudden drop in the X-ray flux at $\delta t \approx 500$ d can be attributed to a pre-existing accretion disk interacting with the tidal disruption stream; however, this assumes that the X-ray emission at $\delta t \gtrsim 500$ d is due to AGN activity, which we argue against here.

We therefore conclude that the only natural explanation for the X-ray emission in terms of its luminosity and temporal behavior is emission from the forward shock.

\subsection{The Dynamical Evolution of the Outflow}\label{sec:outflow}

From our reanalysis of all the observations presented here and our constraints on $\epsilon_B$ (\S\ref{sec:coolfreq}), we calculate the physical parameters of the outflow using the formulation discussed in  \S\ref{subsec:equip} (see Figure 3). We note that the overall trends for all of the parameters are consistent with the results presented in \citealt{BarniolDuran2013} ($\delta t \approx 5 - 645$ d), with a difference in scaling due to the assumed equipartition fractions. The data at $\delta t\approx 700$ d reveal a transition towards Newtonian expansion as the outflow expands into the ambient medium and decelerates. The total energy begins to plateau to a value of $E_K \approx 4 \times 10^{51}$ erg at $\delta t \approx 300$ d, following an order of magnitude increase between $\delta t \approx 30 - 250$ d (Paper I). The jet energy is roughly a factor of $\sim 2$ times larger than found in previous work (Paper I, Paper II) due to our determination of $\epsilon_B\approx 10^{-3}$. Consistent with earlier work, we find $\Gamma \approx 3$ at $\delta t \lesssim 10$ d and an overall evolution given by $\Gamma \propto t^{-0.2}$ out to $\delta t \approx 2000$ d.  At $\delta t \approx 700$ d, the expansion can be characterized as non-relativistic, with $\Gamma \beta \approx 1$. At $\delta t \approx 2000$ d, we find $\beta \approx 0.3$, suggesting that the outflow has evolved from relativistic to non-relativistic on a timescale of a few years. 

We fit a smoothly broken power law to the time evolution of the radius and find that it evolves as $R\propto t^{0.6}$ at $\delta t \approx 10-100$ d and $R \propto t^{0.2}$ at $\delta t \gtrsim 100$ d. The behavior at $\delta t \lesssim 100$ d agrees with our findings in Paper I, and suggests that the radius at early times increases faster than expected for material expanding into a wind ($\rho\propto R^{-2}$) environment, for which $R \propto t^{0.5}$ \citep{Chevalier2000}, but in agreement with a structured outflow launched with a range of Lorentz factors. This further supports our conclusion based on the increase in the total energy by an order of magnitude between $\delta t \approx 30-250$ d (Paper I).

Due to the low magnetic energy density, we find that the strength of the magnetic field is roughly a factor of $\sim 2$ below the equipartition value ($\epsilon_B = 0.1$), evolving as $B \propto t^{-0.3}$ over the timescale $\delta t \approx 10 - 2000$ d. The magnetic field radial profile evolves as $B \propto R$ over the range $R \approx 0.05 - 1.5$ pc. The overall strength of the magnetic field is consistent with the findings for two non-relativistic TDEs with radio emission, ASSASN-14li and XMMSL1 J0740-85 (\citealt{Alexander2016}; \citealt{Alexander2017}). In an upcoming paper, we will consider the implications of the magnetic field strength and present radio polarization observations.

\subsection{The Radial Density Profile}\label{sec:density}

In Figure 4, we plot the inferred radial density profile around Sw 1644+57. Due to our determination that $\epsilon_B\approx 10^{-3}$, we find that the density is roughly a factor of 2 times lower than previously inferred, with $n_{\rm ext} \approx 0.05 - 2 \rm \ cm^{-3}$ at $R \approx 1.5 - 0.1$ pc, respectively, but that the density profile is consistent with our previous findings (Paper I and II). The density profile at $R \approx 0.1 - 0.4$ pc is roughly consistent with $n_{ext} \propto R^{-3/2}$ (though steeper below $R \approx 0.1$ pc), followed by a flattening at $R \approx 0.4 - 0.7$ and a further steepening back to $R^{-3/2}$ out to $R \approx 1.5$ pc. The uniform density at $R \approx 0.4 - 0.6$ is consistent with an expected density enhancement due to Bondi accretion, below which $n_{ext} \propto R^{-3/2}$ \citep{Bondi1952}, as described in Paper I. Outside the Bondi radius, however, the density profile is expected to remain constant. The observed decline may be indicative of mass loss from stellar winds, as is believed to be the case for the Galactic center (\citealt{Melia1992}; \citealt{Baganoff2003}; \citealt{Quataert2004}). In Figure 4, we also plot a model of the Galactic center density profile, in which stellar winds from massive stars have been argued to supply the bulk of the gas content in the central parsec \citep{Quataert2004}. In this model, beyond $R\approx 0.4$, the density profile scales as $n_{ext} \propto R^{-2}$. In the case of Sw 1644+57, the steepening in the density profile at $R \approx 0.6$ pc may therefore reflect a density profile driven primarily by mass loss from massive stars. 

Also shown in Figure 4 is the density profile for the Galactic center, as inferred from X-ray observations \citep{Baganoff2003}, M87 \citep{Russell2015}, and the density profiles in the circumnuclear regions around two non-relativistic TDEs, ASASSN-14li \citep{Alexander2016} and XMMSL1 J0740-85 \citep{Alexander2017}. In the case of the TDEs, the inferred density profiles are computed for both spherical and collimated outflows. For comparison, we scale the radii by the Schwarzschild radius for each black hole, assuming $M_{\rm BH} = 10^{6.5} \rm \  M_{\odot}$ in the case of Sw 1644+57. We find that the density around Sw 1644+57 is significantly lower than the densities inferred for ASASSN-14li and XMMSL1 J0740-85, as well as the Galactic center. On the other hand, it is comparable to the average density in the circumnuclear region of M87.

\section{Implications for Off-Axis TDEs}\label{sec:offaxis}

In Figure 7, we plot the radio light curves of Sw 1644+57 at K, C, and L band ($\sim 22$, $\sim 6$, and $\sim 1.5$ GHz, respectively) extending to $\delta t \approx 1894$ d. At $\delta t \approx 500$ d, the emission at both K and C bands begins to decline and steepen, as $\nu_m$ passes through the bands. Extrapolating the light curves forward in time and using a 3$\sigma$ sensitivity for the VLA in a single hour observation, we find that the emission from Sw 1644+57 will continue to be detectable for $\delta t \approx 15 - 100$ yr at K and C band, respectively. The construction of a more sensitive next generation-VLA would extend the detectability of Sw 1644+57 to several centuries. Finally, we note that Sw 1644+57 may still be detectable with the Very Long Baseline Array (VLBA) at 22 GHz until $\delta t \approx 8$ yr. However, assuming that the jet maintains its collimation, the projected radius will be well below the best-case VLBA angular resolution of $\approx 0.2$ mas (FWHM). It is therefore unlikely that the source can be resolved.

On-axis events such as Sw 1644+57 likely represent a small fraction of the total TDE population ($\lesssim 1\%$; \citealt{Stone2016}). In the case of off-axis TDEs, radio emission is observable at late times ($\delta t \approx$ months - years) as the outflow decelerates and spreads (\citealt{vanVelzen2013}; \citealt{Metzger2015}; \citealt{Stone2016}).  An off-axis event similar to Sw 1644+57 would become detectable at $\delta t \approx 1-2$ yr (the point at which the outflow becomes non-relativistic), but observations spanning a comparable timescale will be required in order to detect significant variation at frequencies of $\gtrsim 5$ GHz. The slow evolution observed at 1.5 GHz would inhibit the recognition of such an event as a transient. Following a peak in the L band light curve at $\delta t \approx 1$ yr, the flux density drops by less than a factor of 2 over a timescale $\delta t \approx 5$ yr. Furthermore, the effects of interstellar scintillation which are prominent at low frequencies may result in amplitude modulations of this order. At the low frequencies of the VLA Sky Survey ($2-4$ GHz), identifying these events as transient sources will therefore be particularly challenging. Thus, while the sensitivity of existing radio facilities allows for the detection of off-axis events with similar properties to Sw 1644+57 to $z\approx 2$ at $\nu \approx 22$ GHz, it is likely that these events will not be easily recognizable as transient sources in a survey spanning only a few years (see also \citealt{Metzger2015}).

\section{Conclusion}\label{sec:conc}

We presented new radio and X-ray observations of Sw 1644+57 extending to $\delta t \approx 2000$ d. We combined these data with optical/NIR measurements at $\delta t\approx 10-200$ d \citep{Levan2016}, and undertook a more robust modeling and fitting approach. Our analysis reveals several new insights:

\begin{itemize}
\item{We find $\epsilon_B \approx 10^{-3}$, suggesting that the source is not in equipartition (for $\epsilon_e \approx 0.1$). We can furthermore rule out a time variable deviation from equipartition based on the excellent fit to the optical/NIR and X-ray light curves over two orders of magnitude in time based on an extrapolation of the radio SEDs with fixed equipartition values.}

\item{The X-ray emission beginning at $\delta t \approx 500$ d and extending to $\delta t \approx 2000$ d can be explained by emission from the forward shock. In analogy with X-ray binaries which exhibit distinct modes of accretion (\citealt{Shakura1973}; \citealt{Novikov1973}), continued X-ray monitoring of Sw 1644+57 may catch the jet turning back on (\citealt{Strubbe2009}; \citealt{Tchekhovskoy2014}), as the accretion rate nears a few percent of Eddington at $\delta t \approx 5000$ d.}
\smallskip

\item{The kinetic energy of the outflow increases to a plateau value of $E_{K} \approx 4 \times 10^{51}$ erg at $\delta t \gtrsim 300$ d. The radius evolves as $R \propto t^{0.6}$ until $\delta t \approx 100$ d, followed by a flattening to $R \propto t^{0.2}$. The rapid increase in energy between $\delta t \approx 30-250$ d and the temporal evolution of the radius at $\delta t \lesssim 100$ d are both indicative of a velocity structure in the outflow.}

\item{The outflow transitions to Newtonian expansion ($\Gamma \beta \approx 1$) at $\delta t \approx 700$ d, with an overall evolution of $\Gamma \propto t^{-0.2}$, tracing the complete evolution of the jet in both the relativistic and non-relativistic regimes.}

\item{The density profile follows $n_{ext} \propto R^{-3/2}$ between $R \approx 0.1 - 0.4$, followed by a flattening at $R \approx 0.4 - 0.7$ and a subsequent steepening back to $n_{ext} \propto R^{-3/2}$ to $R \approx 1.5$ pc. The overall scales are consistent with Bondi accretion and suggest that the bulk of the gas within the central parsec is supplied by winds from massive stars.}
\end{itemize}

Projecting forward, we expect the X-ray emission from Sw 1644+57 to continue tracking the forward shock in tandem with the radio emission. Continued X-ray monitoring may reveal state transitions of the accretion mode, allowing for a determination of the accretion rate at that time, and providing a useful metric to assess the relationship between the relativistic outflow and the accretion mechanism. Radio observations will continue to track the density profile out to $R \approx 2.5$ pc, with the expectation that it will continue to evolve as $n_{ext} \propto R^{-3/2}$. 

Although on-axis events similar to Sw 1644+57 likely represent a small fraction of the total TDE population, radio emission from off-axis events should be more ubiquitous. Existing radio facilities will be capable of detecting these events out to $z \sim 2$, with upcoming facilities extending the range of detectability further, however, the long timescales in the radio suggest that the characterization of these events as transients will be particularly challenging. 

\acknowledgments \textit{Acknowledgments.} We thank Rodolfo Barniol Duran and James Guillochon for their valuable feedback on this work. The Berger Time-Domain Group is supported in part by NSF grant AST-1411763 and NASA ADA grant NNX15AE50G. Support for BAZ is based in part while serving at the NSF. This research has made use of NASA's Astrophysics Data System. 
\software{CASA \citep{McMullin2007}, CIAO (v4.7; \citealt{Fruscione2006}), emcee \citep{ForemanMackey2013}, pwkit \citep{Williams2017},  XSPEC \citep{Arnaud1996}}.

\bibliography{ref}
\bibliographystyle{apj}

\begin{deluxetable*}{ccc}
\tablecolumns{3}
\tablewidth{0pt} 
\tablecaption{Karl G.~Jansky Very Large Array Radio Observations of Sw 1644+57. }
\tablehead{
	\colhead{$\delta t$} & 
	\colhead{Frequency} & 
	\colhead{Flux Density} \\ 
	\colhead{[d]} &
	\colhead{[GHz]} &
	\colhead{[mJy]}
}
\startdata
1105.0&1.3&1.85$\pm$0.12 \\		
1894.0&1.4&1.94$\pm$0.11 \\		
1032.0&1.6&3.27$\pm$0.10 \\	
1105.0&1.6&2.85$\pm$0.12 \\		
1373.0&1.6&1.93$\pm$0.10 \\		
1894.0&1.6&2.17$\pm$0.07 \\		
787.6&1.8&2.40$\pm$0.09 \\	
651.1&2.6&7.98$\pm$0.05 \\	
1105.0&2.6&4.05$\pm$0.08 \\		
1373.0&2.6&2.76$\pm$0.19 \\		
1894.0&2.6&2.08$\pm$0.03 \\		
651.1&3.4&8.78$\pm$0.04 \\	
1105.0&3.4&3.97$\pm$0.08 \\		
1373.0&3.4&2.65$\pm$0.18 \\		
1894.0&3.4&1.88$\pm$0.02 \\		
645.0&4.9&8.24$\pm$0.05 \\
651.1&4.9&8.63$\pm$0.05 \\	
787.6&4.9&6.23$\pm$0.18 \\	
1032.0&4.9&4.21$\pm$0.02 \\	
1105.0&4.9&3.52$\pm$0.05 \\	
1373.0&4.9&2.34$\pm$0.05 \\		
1894.0&4.9&1.47$\pm$0.02 \\		
645.0&7.1&6.91$\pm$0.06 \\
787.6&7.1&5.18$\pm$0.16 \\	
651.1&7.1&7.19$\pm$0.05 \\	
1032.0&7.1&3.43$\pm$0.03 \\		
1105.0&7.1&2.64$\pm$0.06 \\	
1373.0&7.1&1.82$\pm$0.04 \\		
1894.0&7.1&1.15$\pm$0.02 \\		
651.1&8.5&5.99$\pm$0.06 \\	
1029.0&8.5&3.13$\pm$0.02 \\		
1373.0&8.5&1.48$\pm$0.05 \\		
1894.0&8.5&0.93$\pm$0.02 \\		
651.1&10.9&5.05$\pm$0.06 \\	
1029.0&10.9&2.56$\pm$0.03 \\				
1351.0&10.9&1.20$\pm$0.06 \\		
1894.0&10.9&0.76$\pm$0.03 \\							
1029.0&13.5&2.05$\pm$0.03 \\				
1029.0&14.5&1.92$\pm$0.02 \\		
639.0&19.2&3.17$\pm$0.03 \\
787.6&19.2&2.40$\pm$0.03 \\	
1029.0&19.2&1.44$\pm$0.02 \\				
1093.0&19.2&1.16$\pm$0.04 \\		
1351.0&19.2&0.73$\pm$0.04 \\		
1890.0&19.2&0.45$\pm$0.05 \\		
639.0&24.5&2.52$\pm$0.03 \\
787.6&24.5&1.88$\pm$0.03 \\	
1029.0&24.5&1.15$\pm$0.02 \\			
1093.0&24.5&0.92$\pm$0.07 \\	
1890.0&24.5&0.37$\pm$0.06 \\		
639.0&30&1.88$\pm$0.09 \\
639.0&37&1.53$\pm$0.11 \\		
\enddata
\tablecomments{Data prior to $\delta t =$ 639 d are given in Papers I $\&$ II (\citealt{Berger2012}; \citealt{Zauderer2013}).}
\end{deluxetable*}

\begin{deluxetable*}{cccccc}
\tablecolumns{6}
\tablecaption{\textit{Chandra} X-ray Observations of Sw 1644+57. }
\tablehead{
	\colhead{$\delta t$} &
	\colhead{Exposure Time} &
	\colhead{Net Counts} &
	\colhead{Net Count Rate} &
	\colhead{Source Significance }&
	\colhead{Flux ($0.3 - 10$ keV)} \\
	\colhead{[d]} &
	\colhead{[ks]} &
	\colhead{} &
	\colhead{[cts/s]} &
	\colhead{[$\sigma$]} &
	\colhead{[ergs/cm$^2$/s]}	
}
\startdata
612 & 24.7&6.6 $\pm$ 2.5 &(2.6 $\pm$ 1.0) $\times10^{-4}$&3.2& 7.01 $\substack{+5.82 \\ -3.72} \times 10^{-15}$ \\
1425 & 27.8&12.9 $\pm$ 3.6 &(4.6 $\pm$ 1.3) $\times10^{-4}$&6.6& 1.51 $\substack{+0.87 \\ -0.67} \times 10^{-14}$\\
1473 & 18.7&3.9 $\pm$ 2.0 & (2.1 $\pm$ 1.0) $\times10^{-4}$& 2.1 & 6.59 $\substack{+7.14 \\ -4.23} \times 10^{-15}$\\
1611 & 19.7&0 &< 4.1 $\times10^{-4}$ & --  & < 5.59 $\times 10^{-15}$ \\ 
1771 & 26.7&6.4 $\pm$ 2.4& (2.4 $\pm$ 1.0) $\times10^{-4}$& 3.0 &  6.97 $\substack{+5.95 \\ -3.84}\times 10^{-15}$\\
1836 & 24.6&5.8 $\pm$ 2.4 & (2.3 $\pm$ 1.0) $\times10^{-4}$& 2.9 & 6.46 $\substack{+5.93 \\ -3.66} \times 10^{-15}$\\
1910 & 24.6&4.8 $\pm$ 2.1 & (1.9 $\pm$ 0.9) $\times10^{-4}$& 2.4 & 1.11 $\substack{+1.43 \\ -0.81}  \times 10^{-14}$\\
2019 & 24.6&2.6 $\pm$ 1.5 & (1.1 $\pm$ 0.6) $\times 10^{-4}$ & 1.3 & 5.00 $\substack{+8.50 \\ -3.67} \times 10^{-15}$\\
\enddata
\tablecomments{Observed X-ray counts and source significance (1$\sigma$) for Sw 1644+57. Unabsorbed X-ray fluxes are calculated assuming an absorbed power law model with $\Gamma_X = 2.25$, $N_{\rm H,int} = 2 \times 10^{22} \ \rm cm^{-2}$, and $N_{\rm H,MW} = 1.7 \times 10^{20} \ \rm cm^{-2}$. Upper limits correspond to 3$\sigma$ limits. Data for $\delta t = 612$ and $\delta t = 1425,1473$ are from \citet{Zauderer2013} (PI: Tanvir) and \citet{Levan2016} (PI: Levan), respectively.}
\end{deluxetable*}

\begin{deluxetable*}{ccccc}
\tablecolumns{5}
\tablecaption{Radio Spectral Energy Distribution Fits}
\tablehead{
	\colhead{$\delta t$} &
	\colhead{$F_{\nu}$} &
	\colhead{log($\nu_a$)} &
	\colhead{log($\nu_m$)} &
	\colhead{log($\nu_c$)} \\
	\colhead{[d]} &
	\colhead{[mJy]} &
	\colhead{[Hz]} &
	\colhead{[Hz]} &
	\colhead{[Hz]} 
}
\startdata
5&$23.11\substack{+33.07 \\ -9.36}$&$10.71\substack{+0.21 \\ -0.14}$&11.60&$13.22\substack{+0.19 \\ -0.26}$ \\
10&$9.59\substack{+4.35 \\ -1.83}$&$10.11\substack{+0.10 \\ -0.06}$&$11.33 \substack{+ 0.26\\ -0.24}$&$13.86\substack{+0.08 \\ -0.14}$ \\
15&$9.00\substack{+2.87 \\ -1.88}$&$9.98\substack{+0.09 \\ -0.08}$&$11.15\substack{+0.20 \\ -0.17}$&$14.06\substack{+0.09 \\ -0.10}$ \\
22&$12.02\substack{+0.57 \\ -0.78}$&$10.05\substack{+0.01 \\ -0.02}$&$10.76\substack{+0.09 \\ -0.08}$&$14.14\substack{+0.02 \\ -0.01}$ \\
36&$10.79\substack{+1.40 \\ -0.66}$&$9.83\substack{+0.04 \\ -0.03}$&10.85&$14.37\substack{+0.02 \\ -0.03}$ \\
51&$13.87\substack{+2.36 \\ -1.83}$&$9.91\substack{+0.05 \\ -0.06}$&10.84&$14.40\substack{+0.05 \\ -0.05}$ \\
68&$20.36\substack{+1.08 \\ -1.34}$&$9.95\substack{+0.01 \\ -0.02}$&10.71&$14.37\substack{+0.03 \\ -0.02}$ \\
97&$28.26\substack{+2.85 \\ -2.66}$&$9.90\substack{+0.03 \\ -0.03}$&$10.49\substack{+0.06 \\ -0.05}$&$14.40\substack{+0.03\\ -0.03}$ \\
126&$32.14\substack{+1.92 \\ -2.90}$&$9.91\substack{+0.02 \\ -0.03}$&$10.33\substack{+0.03 \\ -0.03}$&$14.49\substack{+0.02 \\ -0.01}$ \\
161&$39.06\substack{+9.54 \\ -6.44}$&$9.86\substack{+0.04 \\ -0.04}$&$10.11\substack{+0.06 \\ -0.06}$&$14.58\substack{+0.03 \\ -0.03}$ \\
197&$33.65\substack{+5.54 \\ -5.49}$&$9.78\substack{+0.04 \\ -0.04}$&$10.06\substack{+0.06 \\ -0.04}$&$14.70\substack{+0.04 \\ -0.03}$ \\
216&$40.58\substack{+5.65 \\ -4.12}$&$9.86\substack{+0.02 \\ -0.02}$&$10.07\substack{+0.03 \\ -0.03}$&$14.67\substack{+0.02 \\ -0.02}$ \\
244&$36.07\substack{+4.97 \\ -2.74}$&$9.84\substack{+0.02 \\ -0.02}$&$10.07\substack{+0.02 \\ -0.04}$&$14.73\substack{+0.02\\ -0.02}$ \\
301&$34.71\substack{+3.30 \\ -15.09}$&$9.81\substack{+0.02 \\ -0.03}$&$9.81\substack{+0.07 \\ -0.14}$&$14.84\substack{+0.02 \\ -0.01}$ \\
390&$25.77\substack{+2.57\\ -7.95}$&$9.58\substack{+0.01\\ -0.02}$&$9.55\substack{+0.04 \\ -0.09}$&$14.94\substack{+0.02 \\ -0.02}$ \\
457&$17.18\substack{+5.34 \\ -9.30}$&$9.54\substack{+0.02 \\ -0.02}$&$9.47\substack{+0.10 \\ -0.14}$&$15.03\substack{+0.03 \\ -0.04}$ \\
582&$5.43\substack{+3.08 \\ -3.24}$&$9.50\substack{+0.02 \\ -0.02}$&$9.29\substack{+0.08 \\ -0.15}$&$15.11\substack{+0.03 \\ -0.04}$ \\
645&$5.93\substack{+2.81 \\ -3.38}$&$9.31\substack{+0.04 \\ -0.05}$&$9.09\substack{+0.09 \\ -0.15}$&$15.28\substack{+0.03 \\ -0.04}$ \\
791&$1.73\substack{+1.29 \\ -1.03}$&$9.43\substack{+0.03 \\ -0.07}$&$9.09\substack{+0.07 \\ -0.16}$&$15.19\substack{+0.11 \\ -0.06}$ \\
1030&$1.32\substack{+0.74 \\ -0.78}$&$9.27\substack{+0.03 \\ -0.03}$&$8.92\substack{+0.06 \\ -0.15}$&$15.39\substack{+0.05 \\ -0.05}$ \\
1100&$1.19\substack{+0.51 \\ -0.57}$&$9.26\substack{+0.02 \\ -0.02}$&$8.92\substack{+0.06 \\ -0.12}$&$15.42\substack{+0.04 \\ -0.04}$ \\
1362&$0.46\substack{+0.23 \\ -0.28}$&$9.24\substack{+0.02 \\ -0.02}$&$8.81\substack{+0.07 \\ -0.16}$&$15.54\substack{+0.07 \\ -0.07}$ \\
1894&$0.28\substack{+0.11 \\ -0.13}$&$9.07\substack{+0.02 \\ -0.02}$&$8.59\substack{+0.06 \\ -0.10}$&$16.00\substack{+0.05 \\ -0.09}$ \\	
\enddata
\tablecomments{The model is described in \S\ref{sec:model} and in \citet{Granot2002}. Values for $\nu_m$ that lack uncertainties correspond to lower limits.}
\end{deluxetable*}

\begin{deluxetable*}{ccccccc}
\tablecolumns{7}
\tablewidth{0pt} 
\tablecaption{Inferred parameters of the relativistic outflow and environment of Sw 1644+57 from model fits of the individual multi-frequency epochs shown in Figure 1. }
\tablehead{
	\colhead{$\delta t$} &
	\colhead{log($E_{K}$)} &
	\colhead{log($R$)} &
	\colhead{log($B$)} &
	\colhead{$\Gamma$} & 
	\colhead{log($N_e$)} & 
	\colhead{log($n_{\rm ext}$)}  \\
	\colhead{[d]} &
	\colhead{ [erg]} &
	\colhead{[cm]} &
	\colhead{[G]} &
	\colhead{ } &
	\colhead{ } &
	\colhead{[cm$^{-3}$]}
}
\startdata
5 & $50.27 \substack{+0.11\\ -0.09}$ & $17.33 \substack{+0.04\\-0.05} $& $ -0.45\substack{+0.09\\-0.07}$ &  $ 3.45\substack{+0.15\\-0.17}$ & $ 51.86\substack{+0.15\\-0.11}$ & $ 0.77\substack{+0.23\\-0.17}$\\
10 &$ 50.48 \substack{+0.08\\ -0.05}$ & $17.69  \substack{+0.03\\-0.04}$ & $ -0.89\substack{+0.09\\-0.06}$ & $ 3.69 \substack{+0.12\\-0.15}$ & $ 51.96\substack{+0.07\\-0.04}$ & $ -0.21\substack{+0.13\\-0.09}$\\
15 & $50.61 \substack{+0.06\\ -0.04 }$ & $17.82  \substack{+0.02\\-0.03} $& $ -1.03\substack{+0.06\\-0.04}$ & $ 3.52\substack{+0.09\\-0.11}$ & $ 52.12\substack{+0.05\\-0.04}$ & $ -0.45\substack{+0.10\\-0.09}$\\
22 &  $50.63 \substack{+ 0.03\\ -0.03 }$ & $17.88  \substack{+0.01\\-0.01}$ & $ -1.10\substack{+0.02\\-0.02}$ & $ 3.14\substack{+0.03\\-0.03}$ & $ 52.30\substack{+0.01\\-0.01}$ & $ -0.45\substack{+0.03\\-0.03}$\\
36 &  $50.93 \substack{+0.05 \\ -0.08 }$ & $18.06  \substack{+0.02\\-0.02}$ & $ -1.22\substack{+0.05\\-0.06}$ & $ 3.02\substack{+0.06\\-0.06}$ & $ 52.53\substack{+0.02\\-0.01}$ & $ -0.75\substack{+0.07\\-0.07}$\\ 
51 &  $51.08 \substack{+0.06 \\ -0.08 }$ & $18.07  \substack{+0.03\\-0.03}$ & $ -1.16\substack{+0.06\\-0.07}$ & $ 2.60\substack{+0.08\\-0.07}$ & $ 52.74\substack{+0.03\\-0.03}$ & $ -0.57\substack{+0.09\\-0.11}$\\
68 &  $51.23 \substack{+0.03 \\ -0.04 }$ & $18.13  \substack{+0.01\\-0.01}$ & $ -1.18\substack{+0.02\\-0.03}$ & $ 2.45\substack{+0.03\\-0.02}$ & $ 52.95\substack{+0.01\\-0.02}$ & $ -0.55\substack{+0.04\\-0.05}$\\
97 &  $51.40 \substack{+0.02 \\ -0.01 }$ & $18.26  \substack{+0.01\\-0.01}$ & $ -1.28\substack{+0.02\\-0.02}$ & $ 2.37\substack{+0.03\\-0.03}$ & $ 53.16\substack{+0.01\\-0.01}$ & $ -0.70\substack{+0.04\\-0.05}$\\
126 &  $51.43 \substack{+0.01 \\ -0.01 }$ & $18.28  \substack{+0.02\\-0.01}$ & $ -1.30\substack{+0.01\\-0.02}$ & $ 2.17\substack{+0.03\\-0.02}$ & $ 53.26\substack{+0.01\\-0.01}$ & $ -0.68\substack{+0.02\\-0.05}$\\
161 &  $51.50 \substack{+0.01 \\ -0.01}$ & $18.36  \substack{+0.02\\-0.02}$ & $ -1.38\substack{+0.02\\-0.02}$ & $ 2.11\substack{+0.04\\-0.04}$ & $ 53.36\substack{+0.01\\-0.01}$ & $ -0.82\substack{+0.07\\-0.06}$\\
197 &  $51.55 \substack{+0.02 \\ -0.02}$ & $18.42  \substack{+0.02\\-0.02}$ & $ -1.44\substack{+0.03\\-0.03}$ & $ 2.05\substack{+0.05\\-0.04}$ & $ 53.41\substack{+0.01\\-0.01}$ & $ -0.93\substack{+0.07\\-0.08}$\\
216 &  $51.53 \substack{+0.01 \\ -0.01}$ & $18.37  \substack{+0.01\\-0.01}$ & $ -1.36\substack{+0.01\\-0.01}$ & $ 1.89\substack{+0.02\\-0.02}$ & $ 53.46\substack{+0.01\\-0.01}$ & $ -0.74\substack{+0.04\\-0.03}$\\
244 & $51.57 \substack{+0.01 \\ -0.01} $ &  $18.38 \substack{+0.01\\-0.01}$ & $ -1.37\substack{+0.01\\-0.01}$ & $ 1.82\substack{+0.02\\-0.02}$ & $ 53.48\substack{+0.01\\-0.01}$ & $ -0.74\substack{+0.04\\-0.03}$\\
301 &  $51.54 \substack{+0.02 \\ -0.01}$ & $18.39  \substack{+0.02\\-0.02}$ & $ -1.41\substack{+0.02\\-0.02}$ & $ 1.71\substack{+0.03\\-0.02}$ & $ 53.51\substack{+0.01\\-0.01}$ & $ -0.78\substack{+0.04\\-0.06}$\\
390 &  $51.51 \substack{+0.03 \\ -0.02}$ & $18.44 \substack{+0.02\\-0.01}$ & $ -1.50\substack{+0.01\\-0.02}$ &  $ 1.62\substack{+0.02\\-0.02}$ & $ 53.56\substack{+0.02\\-0.01}$ & $ -0.86\substack{+0.03\\-0.04}$\\
457 &  $51.49 \substack{+0.03 \\ -0.03}$ & $18.44  \substack{+0.02\\-0.02}$ & $ -1.51\substack{+0.02\\-0.02}$ & $ 1.54\substack{+0.02\\-0.02}$ & $ 53.55\substack{+0.03\\-0.03}$ & $ -0.87\substack{+0.03\\-0.05}$\\
582 &  $51.44 \substack{+0.03 \\ -0.02}$ & $18.43  \substack{+0.02\\-0.02}$ & $ -1.51\substack{+0.02\\-0.02}$ & $ 1.41\substack{+0.02\\-0.02}$ & $ 53.55\substack{+0.03\\-0.02}$ & $ -0.83\substack{+0.04\\-0.05}$\\
645 &  $51.59 \substack{+0.05 \\ -0.04}$ & $18.62  \substack{+0.06\\-0.04}$ & $ -1.73\substack{+0.05\\-0.06}$ & $ 1.57\substack{+0.07\\-0.05}$ & $ 53.66\substack{+0.04\\-0.03}$ & $ -1.30\substack{+0.11\\-0.14}$\\
791 &  $51.41 \substack{+0.07 \\ -0.04}$ & $18.43  \substack{+0.07\\-0.03}$ & $ -1.54\substack{+0.03\\-0.09}$ & $ 1.30\substack{+0.06\\-0.02}$ & $ 53.56\substack{+0.06\\-0.03}$ & $ -0.84\substack{+0.07\\-0.19}$\\
1030 & $51.48 \substack{+0.04 \\ -0.03}$ & $18.56  \substack{+0.04\\-0.03}$ & $ -1.70\substack{+0.03\\-0.04}$ &  $ 1.32\substack{+0.03\\-0.02}$ & $ 53.63\substack{+0.04\\-0.03}$ & $ -1.15\substack{+0.07\\-0.09}$\\
1100 &  $51.40 \substack{+0.02 \\ -0.02}$ & $18.53  \substack{+0.02\\-0.02}$ & $ -1.69\substack{+0.02\\-0.02}$ & $ 1.27\substack{+0.01\\-0.01}$ & $ 53.57\substack{+0.02\\-0.02}$ & $ -1.12\substack{+0.05\\-0.05}$\\
1362 &  $51.31 \substack{+0.03 \\ -0.02}$ & $18.49  \substack{+0.02\\-0.02}$ & $ -1.66\substack{+0.02\\-0.03}$ & $ 1.14\substack{+0.02\\-0.01}$ & $ 53.54\substack{+0.03\\-0.03}$ & $ -1.02\substack{+0.04\\-0.06}$\\
1894 & $51.49\substack{+0.03 \\ -0.03}$ & $18.63  \substack{+0.02\\-0.01}$ & $ -1.80\substack{+0.01\\-0.03}$ & $ 1.04\substack{+0.01\\-0.01}$ & $ 53.74\substack{+0.03\\-0.04}$ & $ -1.27\substack{+0.03\\-0.08}$\\
\enddata
\tablecomments{Calculations of the inferred parameters are described in \S\ref{sec:model} and in \citet{BarniolDuran2013}.}
\end{deluxetable*}

\begin{figure*}
\centering
\includegraphics[width=500pt]{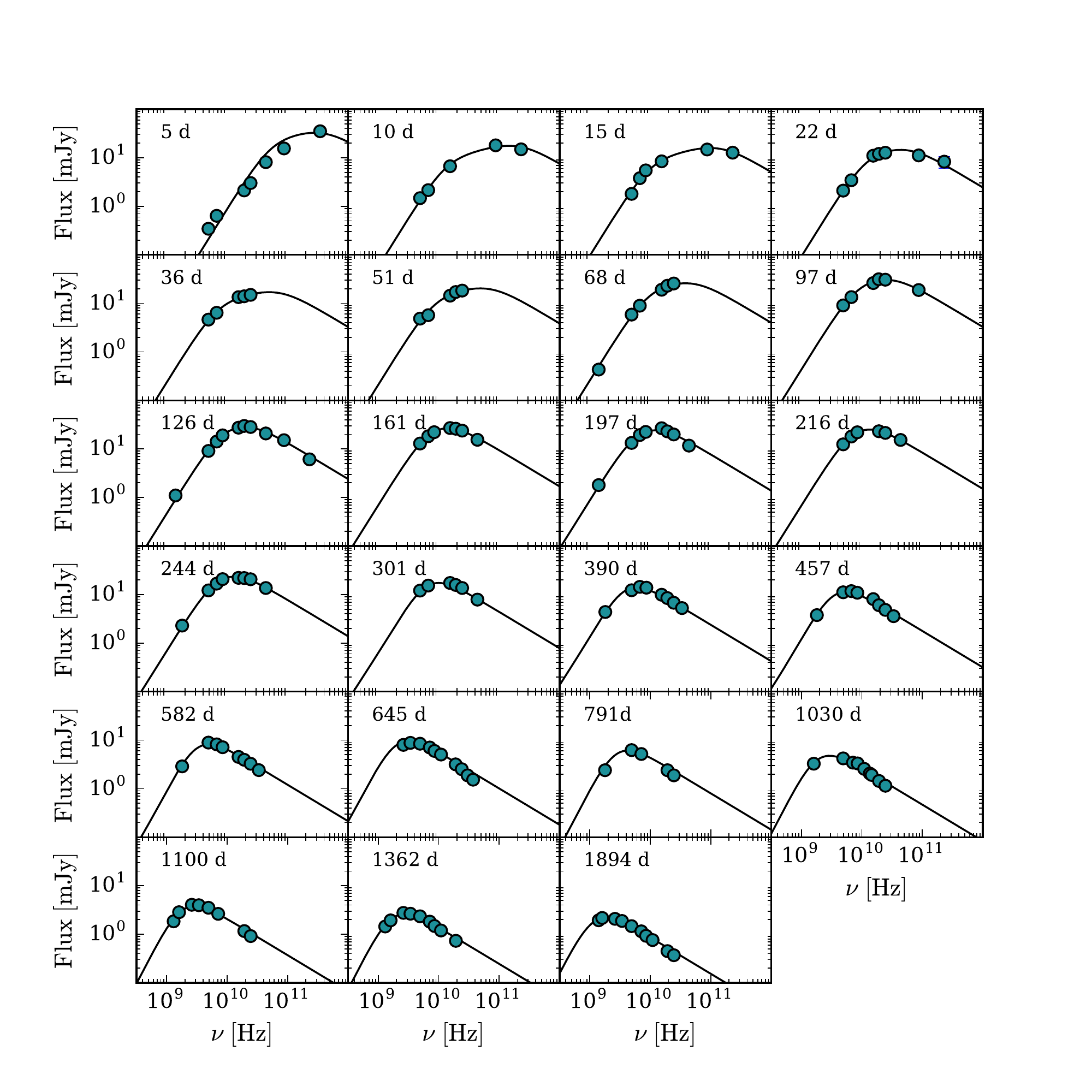}
\caption{Multi-frequency radio spectral energy distributions of Sw 1644+57 at $\delta t \approx 5 - 1894$ d. The solid lines are fits based on the model described in \S\ref{subsec:seds}. In each epoch, we fit for $\nu_a$, $\nu_m$, and $F_{\nu}$  with a fixed value of $p=2.5$. Error bars on individual data points are smaller than the size of the symbols.}
\label{fig:radioseds}
\end{figure*}

\begin{figure*}[ht!]
\centering
\includegraphics[width=500pt]{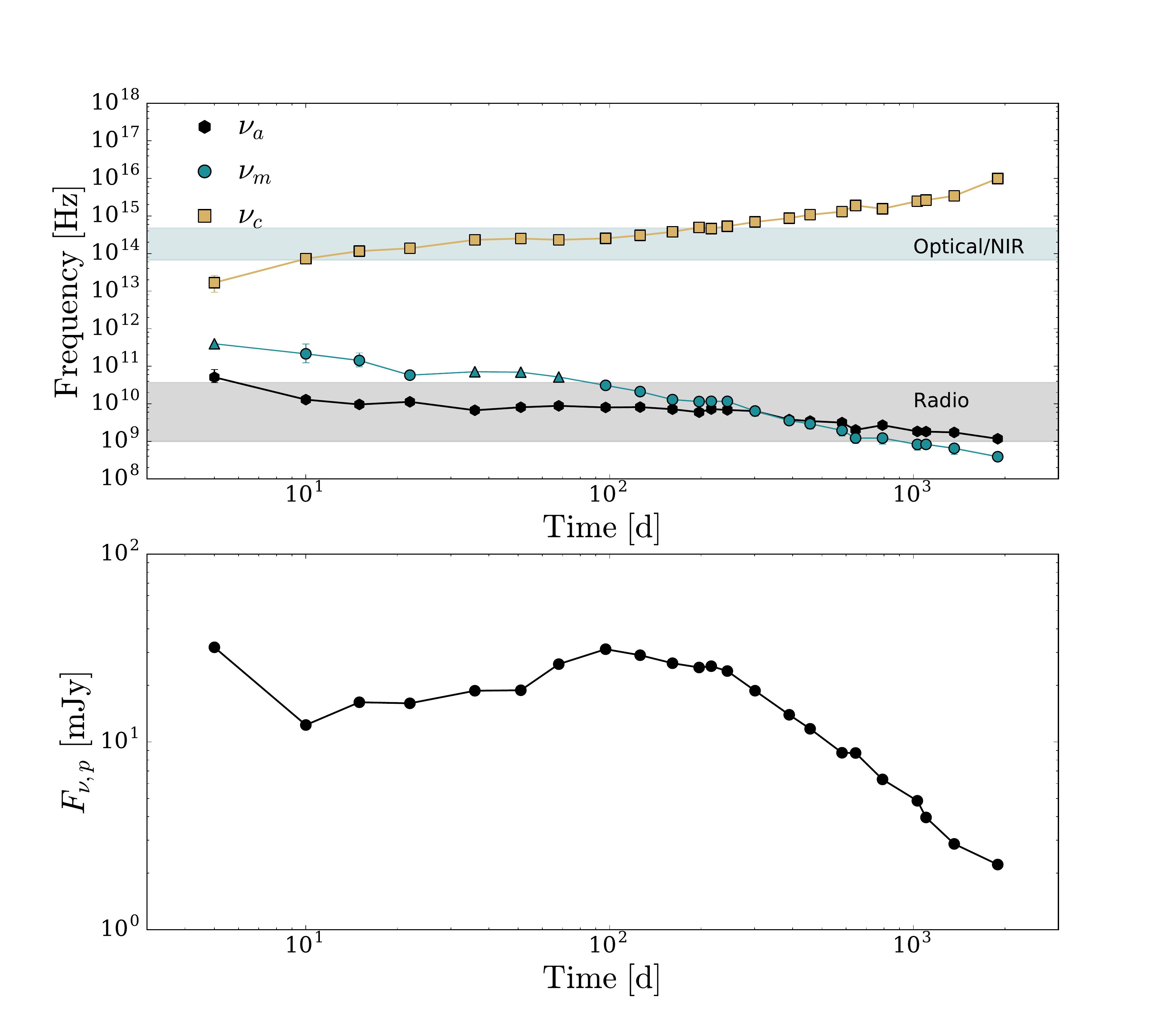}
\caption{\textit{Top:} Temporal evolution of the synchrotron break frequencies $\nu_a$ and $\nu_m$ as derived from fits to the radio SEDs (Figure 1 and \S\ref{subsec:seds}). Triangles indicate lower limits. Also shown is the temporal evolution of the cooling frequency, $\nu_c$ (\S\ref{sec:coolfreq}). \textit{Bottom:} Temporal evolution of $F_{\nu, p}$ as derived from fits to the radio SEDs}
\label{fig:breakfreqs}
\end{figure*}

\begin{figure*}[ht!]
\centerline{\includegraphics[width=600pt]{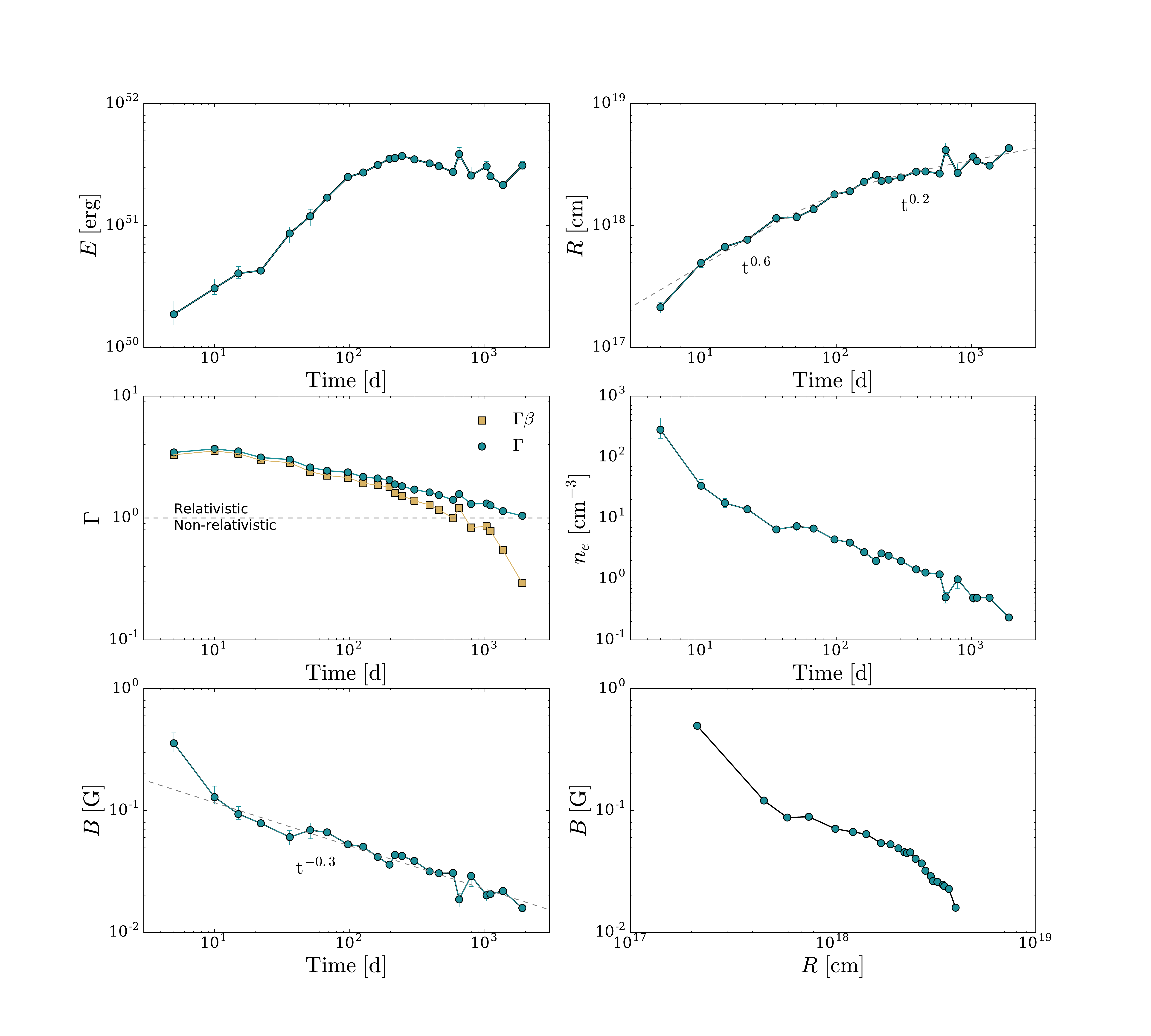}}

\caption{Temporal evolution of the outflow properties as extracted from fits to the SEDs and using the analysis described in \S\ref{subsec:equip}. The beaming-corrected kinetic energy plateaus to a value of $E_K \approx 4 \times 10^{51}$ erg at $\delta t \gtrsim 300$ d, following a rapid increase at earlier times. Consistent with previous work, we find that the radius increases as $R \propto t^{0.6}$ to $\delta t \approx 100$ d, followed by a flattening to $R \propto t^{0.2}$. The Lorentz factor decreases steadily ($\Gamma \propto t^{-0.2}$) from an initial value of $\Gamma\approx 3$, and the outflow becomes non-relativistic ($\Gamma\beta\approx 1$) at $\delta t\sim 700$ d. The value of $n_e$ refers to the number density of radiating particles in the outflow. The magnetic field is roughly a factor of $\sim 2$ below the equipartition value, evolving as $B \propto t^{-0.3}$.}
\end{figure*}

\begin{figure*}[ht!]
\centerline{\includegraphics[width=600pt]{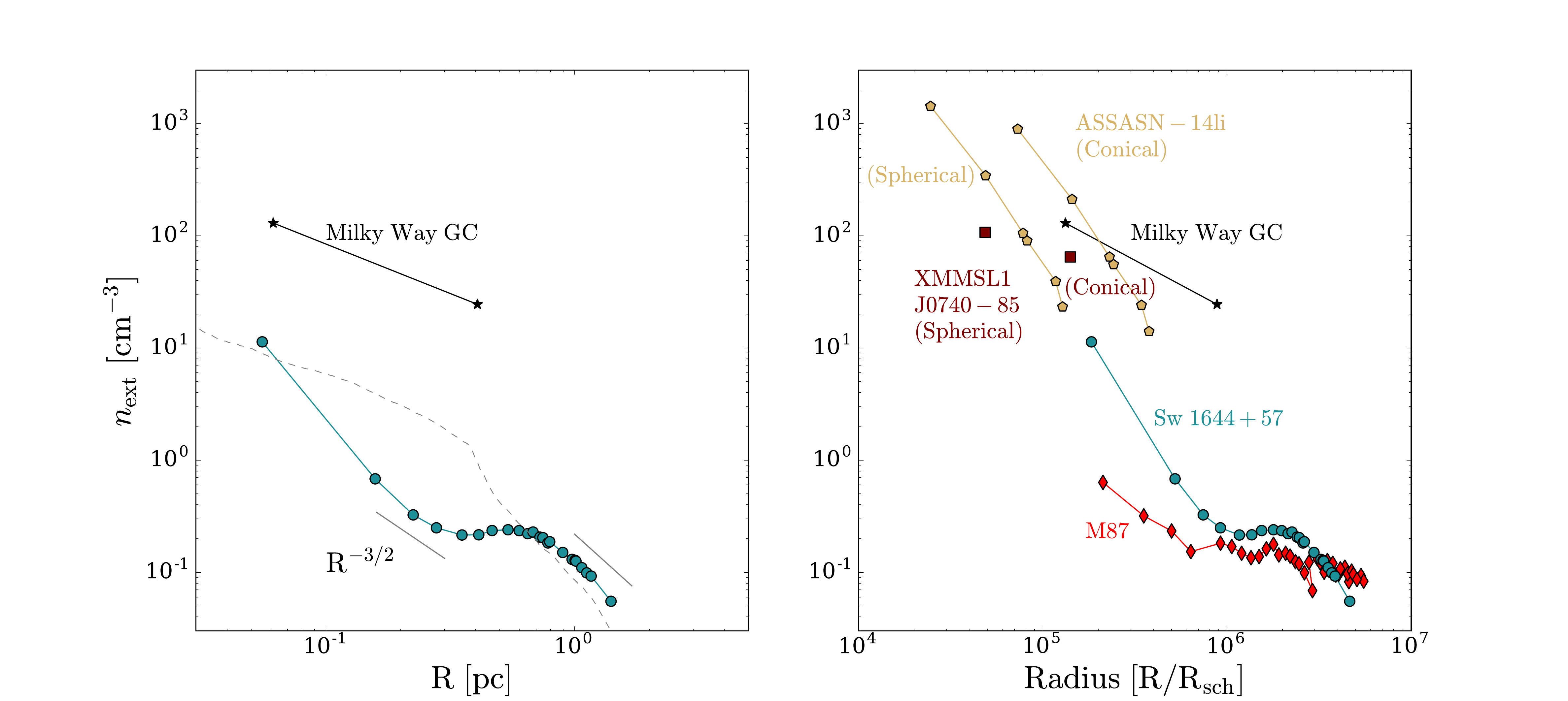}}
\caption{\textit{Left:} Radial density profile around the SMBH as inferred from modeling of the radio SEDs (blue points), compared to the density profile for the Galactic center (black points; \citealt{Baganoff2003}). Also shown as a dashed line is a scaled-down ($n_{\rm ext}/20$) model of the Galactic center density profile in which stellar winds drive the bulk of the gas out of the central parsec  \citep{Quataert2004}. \textit{Right:} The density profile around Sw 1644+57 as compared to the circumnuclear regions of two non-relativistic radio TDEs, ASASSN-14li (yellow; \citealt{Alexander2016}) and XMMSL1 J0740-85 (red; \citealt{Alexander2017}), assuming both spherical and conical outflows, and M87 \citep{Russell2015}. In all cases, the radii are scaled by the Schwarzschild radius of the corresponding SMBH, where we use $M_{\rm BH} = 10^{6.5} \ \rm M_{\odot}$ for Sw 1644+57.}
\end{figure*}

\begin{figure*}[ht!]
\centering
\includegraphics[width=500pt]{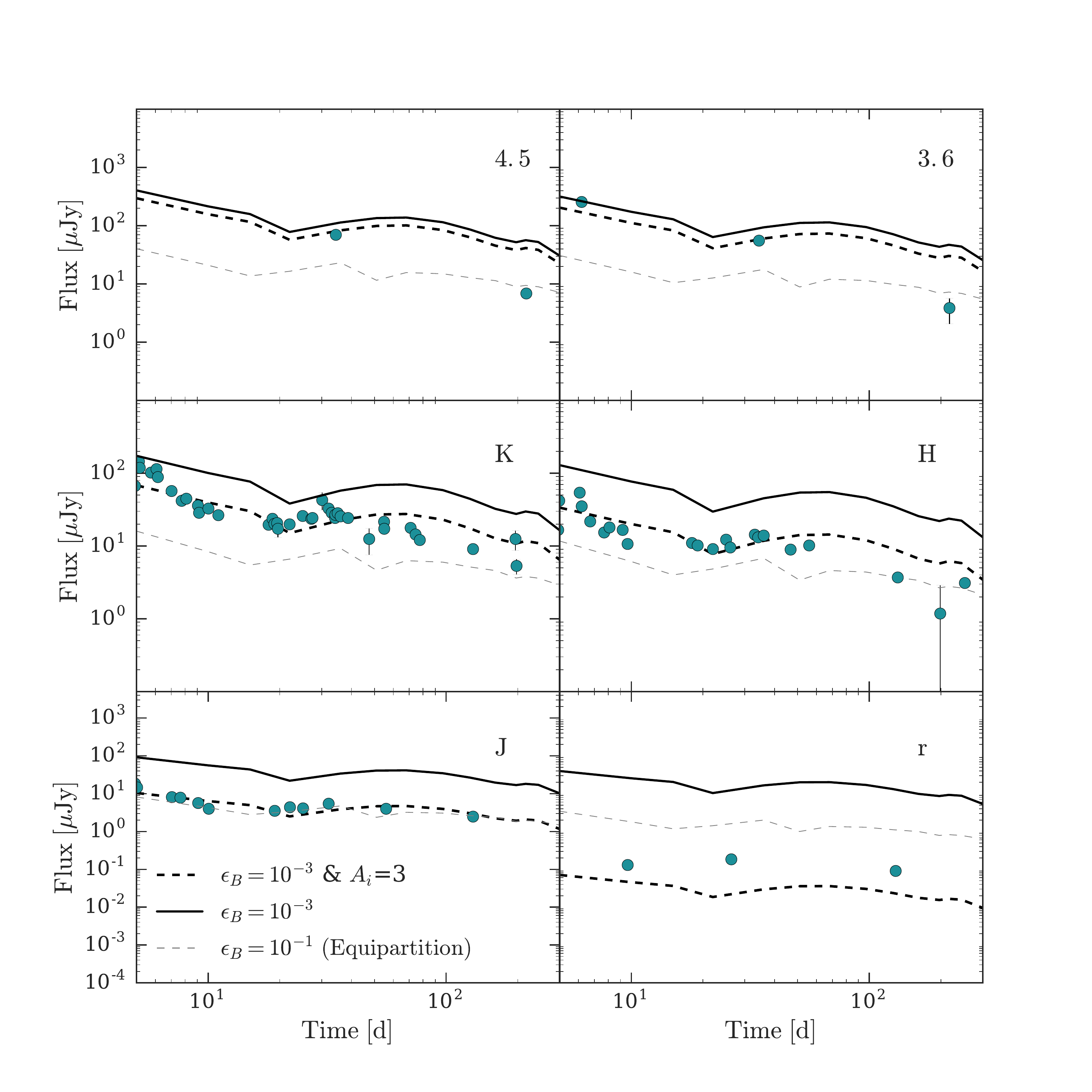}
\caption{Optical/NIR light curves \citep{Levan2016} with models extrapolated from the radio SEDs. The solid black curve illustrates that $\epsilon_B \approx 10^{-3}$ overpredicts the observed optical/NIR flux in the absence of extinction, whereas the forward shock emission can reproduce the observed flux with $A_i = 3$ mag (dashed black curve). On the other hand, an equipartition value of $\epsilon_B \approx 0.1$ (grey dashed curve) under-predicts the measured flux values in several bands and can therefore be ruled out.}
\end{figure*}

\begin{figure*}[ht!]
\centering
\centerline{\includegraphics[width=500pt]{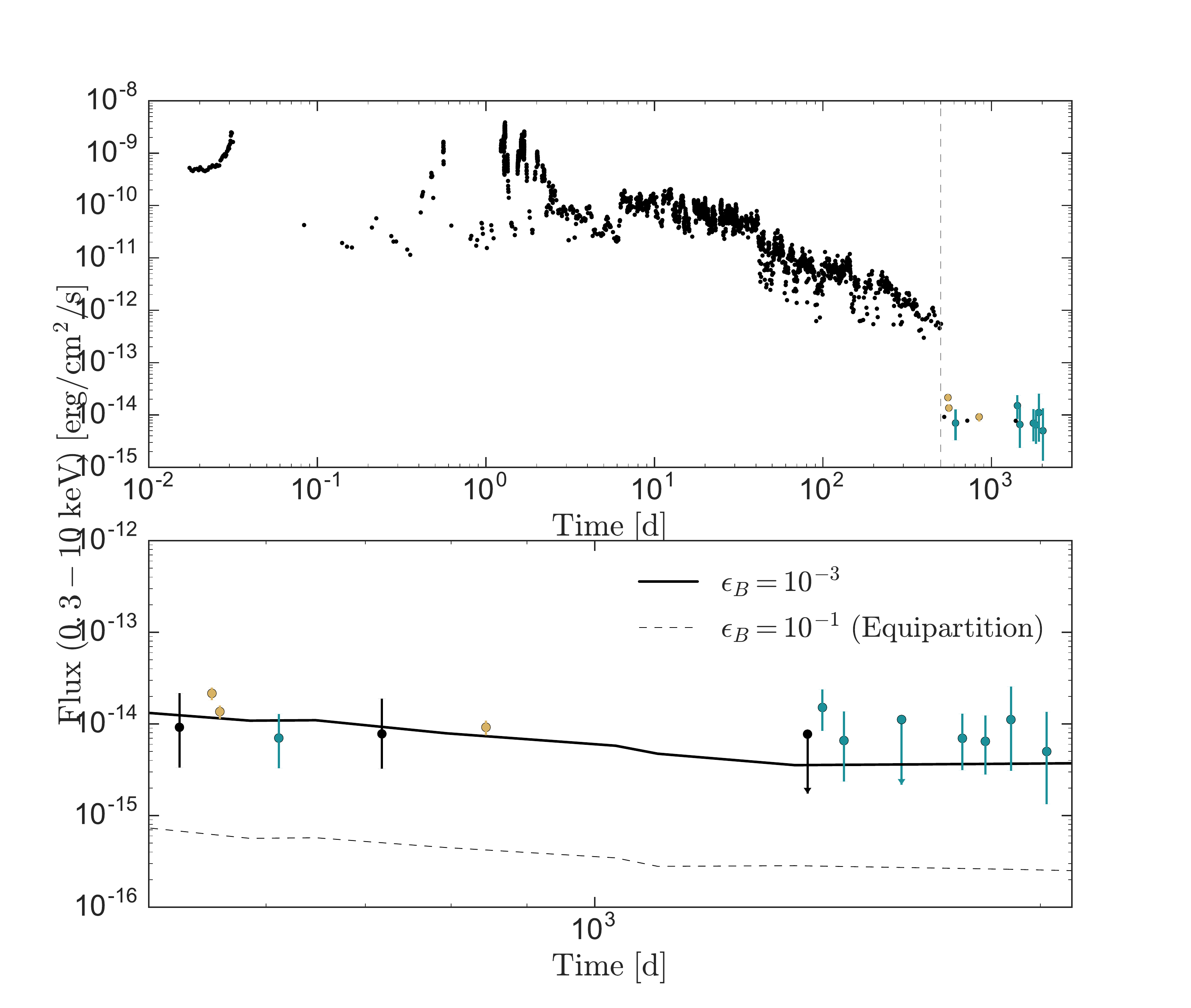}}
\caption{\textit{Top:} X-ray light curve for Sw 1644+57, including data from {\it Swift}/XRT (black points; \citealt{Mangano2016}), XMM-{\it Newton} (yellow; \citealt{Levan2016}) and {\it Chandra} (cyan; see Table 2). The dashed line indicates the transition to forward-shock X-ray emission at $\delta t \approx 500$ d. \textit{Bottom}: The black lines show the X-ray flux from extrapolations of the radio SEDs. While $\epsilon_B \approx 0.1$ (i.e., equipartition) under-predicts the observed X-ray flux, $\epsilon_B \approx 10^{-3}$ is in good agreement with the observed values.}
\end{figure*}

\begin{figure*}[ht!]
\centering
\centerline{\includegraphics[width=400pt]{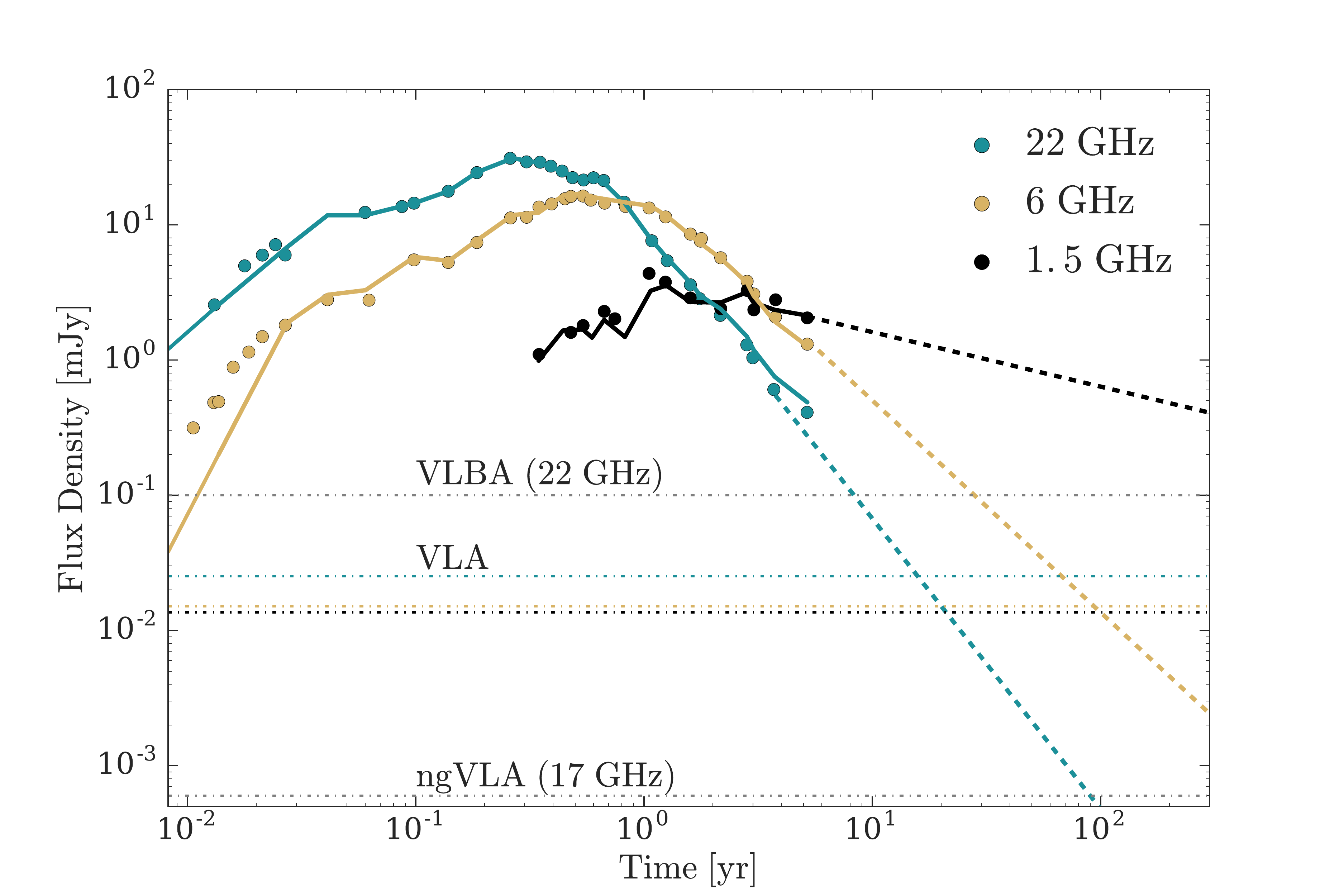}}
\caption{Radio light curves of Sw 1644+57 extending to $\delta t \approx 2000$ d. The solid lines are models based on independent fits of the radio SEDs (Figure 1) using the model described in \S\ref{subsec:seds}. We extrapolate the models forward in time using power-law fits to the late-time epochs (dashed lines). The horizontal lines indicate the 3$\sigma$ sensitivity for the VLA (1 hour observation) and proposed ngVLA (1 hour at 17 GHz), indicating that Sw 1644+57 should be detectable at centimeter wavelengths for decades to centuries. The 3$\sigma$ sensitivity limit for the VLBA suggests that the source may still be detected with very long baseline interferometry until $\delta t \approx 8$ yr, however, the projected angular size at this time will be well below the best-case VLBA angular resolution. }
\end{figure*}

\end{document}